\documentstyle[prl,aps,multicol,epsf]{revtex}
\begin{document}
\title{Why is $d-wave$ pairing in {\em HTS} robust in the presence of impurities?}
\author{{Miodrag L. Kuli\'c}}
\address{Max-Planck-Institut f\"ur Festk\"orperforschung, 70569 Stuttgart, Germany}
\author{{Viktor Oudovenko}}
\address{Max-Planck-Institut f\"ur Festk\"orperforschung, 70569 Stuttgart, Germany\\
Joint Institute for Nuclear Research, 141980 Dubna, Russia}
\date{14 January 1997}
\maketitle

\begin{abstract}
In the recent theory of strong correlations by \cite{Kulic1}~--\cite{Zeyher2}
it was shown that by lowering doping concentration a \underline{forward peak 
} is developed in the charge scattering channel. Accordingly, near the
optimal doping the nonmagnetic scattering is pronounced in the $d$-channel
and $d-wave$ pairing is \underline{robust} against defects and impurities.
It is reflected in a decrease of the slope of the critical temperature $%
T_c(\Gamma _s)$ at $\Gamma _s\rightarrow 0$ by factor $(1-\beta ),$ where $%
\beta $ is the scattering anisotropy parameter and $\Gamma _s$ is an average
scattering rate. For large doping,  $\beta $ is small \cite{Kulic1},\cite
{Zeyher1} and $d-wave$ pairing loses its robustness. 
The theory is
generally formulated for the bi-layer model by including: $(1)$ intra- and
inter-plane pairing; $(2)$ intra- and inter-plane impurities.
\end{abstract}


\begin{multicols}{2}
\narrowtext

\newpage\ 

Recently there were a lot of evidences for $d-wave$ pairing in high-$T_c$%
-superconductors ($HTS$). However, the mechanism of pairing is still
unknown, although many proposals were done. For instance, there were claims
that superconductivity is due to strong correlations\cite{Lee}, or that $AF$
fluctuations are important - see \cite{Scalapino}. Even the electron-phonon
interaction renormalized by strong correlations can give rise to $d-wave$
pairing, as it was shown by Kuli\'c and Zeyher \cite{Kulic1},\cite{Zeyher1},%
\cite{Zeyher2}. Moreover, in\cite{Kulic1},\cite{Zeyher1},\cite{Zeyher2} it
was shown that the charge-charge scattering is renormalized via a
charge-vertex $\gamma (\vec q)$ (screening due to strong correlations) with
a peak in the forward scattering ($\mid \vec q\mid \ll k_F$), while the
backward scattering (large $\mid \vec q\mid $) is suppressed. This means
that a bare nonmagnetic scattering potential $U_0(\vec q)$ is renormalized
by strong correlations $U_0(\vec q)\rightarrow U_{eff}(\vec q)$. Note, in
the Born approximation (after averaging over impurity positions) one has $%
U_{eff}(\vec q)=\gamma ^2(\vec q)u_0^2(\vec q)\equiv u^2(\vec q)$, where $%
u_0(\vec q)$ is the single impurity potential. It follows \cite{Kulic1},\cite
{Zeyher1},\cite{Zeyher2} that when the bare potential is local, i.e. $u_0^2(%
\vec q)=const$, $U_{eff}$  becomes ''long-ranged'' in the presence of strong
correlations. Moreover, within the $t-J$ model \cite{Kulic1},\cite{Zeyher1},%
\cite{Zeyher2} and at some optimal doping $\delta \approx 0.2$ the $d$%
-channel component of $U_{eff}(\vec q)$ is of the same order of magnitude as
the one in the fully symmetric $s$-channel. When $U_0(\vec q)$ is due to the
electron-phonon interaction then superconductivity can be also of $d$-type
in the presence of strong Hubbard repulsion.

If we accept that superconductivity in $HTS$ is of $d-wave$ type, then one
can put a question: why is $d-wave$ pairing robust in the presence of
various kinds impurities and defects in $HTS$ materials, as it was observed
experimentally\cite{Sun1},\cite{Tolpygo}? In this paper we present a simple
theory, which is based on the effective nonmagnetic impurity potential $%
U_{eff}(\vec q)$ with the \underline{large $d$-component} and investigate
its effect on $d-wave$ pairing \cite{Efetov}. Experiments on the irradiation
of samples as well as on substitution of atoms in $HTS$ materials are
analyzed by using the proposed theory. In the following we consider the weak
coupling theory\cite{A-G} and its extension to the \underline{bi-layer}
systems \cite{Hofmann},\cite{Gajic}. In the bi-layer model (adequate for $%
YBa_2Cu_3O_{7-x}$) with the hopping $t_{\perp }$ between two layers
(separated by $Y$ ions) and with the intra- and inter-layer pairings $\Delta
_{\parallel }(\vec p)$ and $\Delta _{\perp }(\vec p)$ respectively the
averaged (over impurity positions) Green's function $\hat G(p)(\equiv \hat G(%
\vec p,i\omega _n))$ is $4\times 4$ matrix. It is given by $\hat G^{-1}(p)=%
\hat G_0^{-1}(p)-\hat \Sigma _{imp}(p)\hat G(p)$, where $\hat G_0(p)$ is the
bare Green's function and the impurity self-energy matrix $\hat \Sigma
_{imp}(p)$ in the bi-layer model is given by\cite{A-G} 
\begin{equation}
\hat \Sigma _{imp}(\vec p,i\omega _n)=c\int \frac{d^2p^{\prime }}{(2\pi )^2}%
\hat U_{eff}^{imp}(\vec p,\vec p^{\prime })\hat G(\vec p^{\prime },i\omega
_n).  \label{eq.1}
\end{equation}
Here, $\omega _n=\pi T(2n+1)$ and $c$ are Matsubara frequencies and the
concentration of impurities respectively. In order to simplify notations we
set $\hat U_{eff}^{imp}(\vec p,\vec p^{\prime })\equiv \hat U(\vec p,\vec p%
^{\prime })$. The structure of the $4\times 4$ matrix $\hat U(\vec p,\vec p%
^{\prime })$ depends on the type of impurities: \underline{Case $I$} - there
are inter-layer impurities (for instance substitution of $Y$ by $\Pr $); 
\underline{Case $II$} - there are intra-layer impurities (for instance
substitution of $Cu$ by $Zn$, or the presence of oxygen vacancies due to ion
damaging etc.). They are explicitly given in \cite{Hofmann} - see also
below. Since, in the weak coupling limit the relevant processes are taking
place near the Fermi surface it is assumed that the matrix $\hat U(\vec p,%
\vec p^{\prime })$ $\equiv \hat U(\theta ,\theta ^{\prime })$ depends on the
angles $\theta ,\theta ^{\prime }$ of momenta on the Fermi surface. In the
following, the definition $\int_0^{2\pi }\frac{d\theta ^{\prime }}{2\pi }%
(...)\equiv \langle (...)\rangle _{\theta ^{\prime }}$ is used. $\hat G^{-1}(%
\vec p,i\omega _n)$ in the bi-layer model is searched in the form analogous
to that given in \cite{Hofmann},\cite{Gajic}, with renormalized Matsubara
frequencies $\tilde \omega _n$ and renormalized superconducting order
parameters $\tilde \Delta _{n,i\text{ }}(\vec p)$ ($i=1,2$ is the band index
-see below)

\begin{equation}
\tilde \omega _n(\theta )=\omega _n+{\frac 12}\langle {\tau _{\bar i\text{ }%
\bar i}^{-1}}(\theta ,\theta ^{\prime })\frac{\tilde \omega _n(\theta
^{\prime })}{\sqrt{\tilde \omega _n^2(\theta ^{\prime })+\tilde \Delta _{n,%
\bar i}^2(\theta ^{\prime })}}\rangle _{\theta ^{\prime }}  \label{eq.2}
\end{equation}
\begin{equation}
\tilde \Delta _{n,i}(\theta )=\Delta _i(\theta )+{\frac 12}\langle {\tau _{i%
\bar j}^{-1}}(\theta ,\theta ^{\prime })\frac{\tilde \Delta _{n,\bar j%
}(\theta ^{\prime })}{\sqrt{\tilde \omega _n^2(\theta ^{\prime })+\tilde 
\Delta _{n,\bar j}^2(\theta ^{\prime })}}\rangle _{\theta ^{\prime }}.
\label{eq.3}
\end{equation}
To this set of equations one should add also the self-consistent gap
equations 
\begin{equation}
\Delta _i(\theta )=T_c\sum_n\langle \lambda _{i\bar j}(\theta ,\theta
^{\prime })\frac{\tilde \Delta _{n,\bar j}(\theta ^{\prime })}{\sqrt{\tilde 
\omega _n^2(\theta ^{\prime })+\tilde \Delta _{n,\bar j}^2(\theta ^{\prime })%
}}\rangle _{\theta ^{\prime }}.  \label{eq.4}
\end{equation}
The summation over the bar indices is assumed everywhere. Here, $\Delta
_i(\theta )$ ($i=1,2$) is the gap in the $i-th$ band, i.e. $\Delta
_{1,2}(\theta )=\Delta _{\parallel }(\vec p)\mp \Delta _{\perp }(\vec p)$
where $\lambda _{ij}(\theta ,\theta ^{\prime })$ ($i,j=1,2$) are the intra-
and inter-band pairing coupling constants\cite{Hofmann}. In the following,
we assume $d-wave$ pairing (instead of the $s-wave$ one used in \cite
{Hofmann}) $\Delta _i(\theta )=\Delta _iY_d(\theta )=-\Delta _i(\theta +\pi
/2)$, $\langle Y_d(\theta )\rangle _\theta =0$, $\langle Y_d^2(\theta
)\rangle _\theta =1,$ and $\lambda _{ij}(\theta ,\theta ^{\prime })=\lambda
_{ij}Y_d(\theta )Y_d(\theta ^{\prime })$. Near the critical temperature $T_{c%
\text{ }}$one can linearize eqs.(2)-(4) which gives 
\begin{equation}
\tilde \omega _n(\theta )=\omega _n[1+{\frac 12}\frac{\langle {\tau _{\bar i%
\bar i}^{-1}}(\theta ,\theta ^{\prime })\rangle _{\theta ^{\prime }}}{\mid
\omega _n\mid }]\equiv \omega _n\eta _n(\theta )  \label{eq.5}
\end{equation}
\begin{equation}
\tilde \Delta _{n,i}(\theta )=\eta _{n,i\bar j}(\theta )\Delta _{\bar j%
}(\theta ),  \label{eq.6}
\end{equation}
where $\tau _{ij}^{-1}(\theta ,\theta ^{\prime })$ and $\eta _{n,i\bar j%
}(\theta )$ depend on the type of impurities (intra- or inter-plane) -see%
\cite{Hofmann}, while $\eta _{n,i\bar j}(\theta )$ are solutions of a
coupled set of integral equations (w.r.t. $\theta $). In the presence of 
\underline{inter-plane} impurities (the case $I$), one has $\tau
_{ij}^{-1}(\theta ,\theta ^{\prime })=\delta _{ij}N_i(\theta )U(\theta
,\theta ^{\prime })$, where $N_i(\theta )$ are the fully symmetric angle
dependent band ($i=1,2$) density of states on the Fermi surface. Note, this
kind of impurities does not affect isotropic $s-wave$ pairing - see \cite
{Hofmann}, but it does affect $d-wave$ pairing \cite{Kulic2}. In the case $II
$ of the \underline{intra-plane} impurities one has $\tau _{ij}^{-1}(\theta
,\theta ^{\prime })=U_i(\theta ,\theta ^{\prime })N_j(\theta )$ with $%
U_1(\theta ,\theta ^{\prime })=U_2(\theta ,\theta ^{\prime })$. Note, the $%
s-wave$ inter-layer pairing is strongly affected by these impurities, due to
the breaking of the local reflection symmetry between layers\cite{Hofmann} -
transitions between symmetric and antisymmetric band occur.

Let us analyze \underline{the case $II$}, which is frequently realized in
the $YBa_2Cu_3O_{7-x}$ family. A very important ingredient of our theory are
strong correlations, for which it has been shown that in $\tau ^{-1}(\theta
,\theta ^{\prime })$ only $s$- and $d$-channels dominate\cite{Kulic1},\cite
{Zeyher1},\cite{Zeyher2}. When generalized to the bi-layer model one has $%
\tau _{ij}^{-1}(\theta ,\theta ^{\prime })=\Gamma _{s,ij}(\theta ,\theta
^{\prime })+\Gamma _{d,ij}Y_d(\theta )Y_d(\theta ^{\prime })+....$ Note, $%
\Gamma _{s,ij}(\theta ,\theta ^{\prime })$($>0)$ is fully symmetric and it
holds $\Gamma _{s,ij}(\theta ,\theta ^{\prime })>\Gamma _{d,ij}Y_d(\theta
)Y_d(\theta ^{\prime })$, where $\Gamma _{d,ij}>0$. Their doping dependence
is analyzed below. Let us assume that the fully symmetric scattering rates
are weakly angle dependent, i.e. $\Gamma _{s,ij}(\theta ,\theta ^{\prime
})\simeq \Gamma _{s,ij}$. In this case eq.(3) is reduced to a set of
algebraic equations for the critical temperature $T_c$ 
\begin{equation}
\Delta _i=T_c\sum_n\lambda _{i\bar j}\frac{\eta _{n,\bar j\bar k}\Delta _{%
\bar k}}{\mid \omega _n\mid \eta _n},  \label{eq.7}
\end{equation}
where $\eta _n=1+(\Gamma _{s,11}+\Gamma _{s,22})/\mid \omega _n\mid $, $\eta
_{n,ij}=\delta _{ij}+\Gamma _{d,jj}/(\mid \omega _n\mid +\Gamma _s-\Gamma _d)
$, $\Gamma _s=\Gamma _{s,11}+\Gamma _{s,22}$, and $\Gamma _d=\Gamma
_{d,11}+\Gamma _{d,22}$. $Eq.(7)$ contains reach physics which depends on
the values and signs of the coupling constants $\lambda _{ij}$ see \cite
{Kulic2}. However, in order to illustrate the basic effect of strong
correlations we simplify the problem by assuming that the inter-layer
pairing is negligible, i.e. $\Delta _{\perp }(\vec p)\approx 0$ and $\Delta
_1\approx \Delta _2\approx \Delta _{\parallel }$. Then $eq.(7)$ is reduced to

\begin{equation}
ln\frac{T_c}{T_c^0}=[\psi (\frac 12)-\psi ((1-\beta )x+\frac 12)],
\label{eq.8}
\end{equation}
where $\beta =(\Gamma _d/\Gamma _s)<1$, $x=$ $\Gamma _s/4\pi T_c$ and $%
T_c^0=1.13\omega _c\exp \{-1/\lambda \}$, $\lambda =\lambda _{11}+\lambda
_{12}$. Similar equation has been obtained also in \cite{Nagi}. It tells us
that the pair-breaking effect of nonmagnetic impurities is strongly reduced
if the scattering potential contains a large anisotropic part which is
compatible with the pairing amplitude. It follows from $eq.(8)$ that in the
presence of an anisotropic scattering ($\beta \neq 0$ ) $d-wave$ pairing is
more \underline{robust} than in the case of the isotropic one ($\beta =0$ ).
This robustness is reflected in the \underline{ decrease of the slope} of $%
T_c(x)$ at $x=0$ by factor $(1-\beta )$. It should be stressed that the
order parameter keeps its $d-wave$ form, i.e. $\Delta (\theta )\approx
\Delta _dY_d(\theta )$ and $\langle \Delta (\theta )\rangle =0$.

The above simple analysis would not deserve to much attention if we would
not be able to find the range of the parameter $\beta $, as well as the
origin for its large value in $HTS$. Luckily, the doping dependence of $%
\beta (\delta )$ can be extracted from the theory of strong correlations 
\cite{Kulic1},\cite{Zeyher1},\cite{Zeyher2}, where for $\delta \leq 0.2$ the
parameter $\beta $ becomes of the order of one(!), while at large doping ($%
1-\delta \rightarrow 0$) one has $\beta \ll 1$ -see $Fig.1$. The latter
value leads to a rapid decrease of $T_c$ by increasing $\Gamma _s$. These
results tell us that $d-wave$ pairing in the underdoped and optimally doped $%
HTS$ materials is robust in the presence of impurities, while it is much
less robust in the overdoped systems - see $Fig.1$.
\\[-0.2cm]
\epsfysize=4.8in
\hspace*{1.0cm}\epsffile{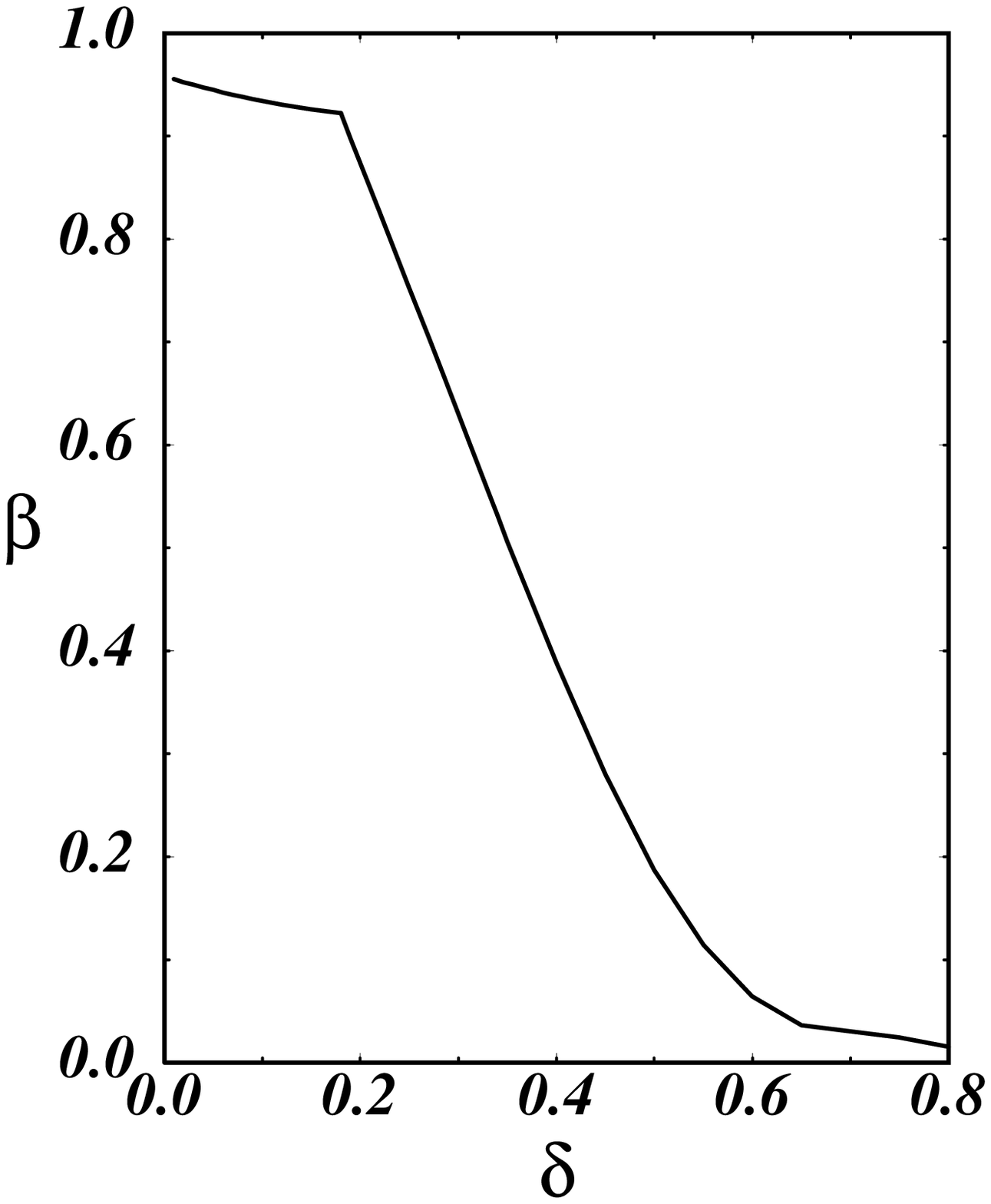}
\vspace*{-2.5cm}
{FIG. 1. The anisotropic scattering parameter $\beta $ as a function of the
doping concentration $\delta $ based on the work \cite{Zeyher1}.}

The average pair-breaking parameter $\Gamma _s$ is extracted from the
measurements of the residual resistivity $\rho _i=4\pi /\tau _{tr}\omega
_{pl}^2$, which is due to the substitutional defects ($Cu\rightarrow Zn$, $%
Y\rightarrow $Pr) or due to the ion ($Ne^{+}$) or electron damaging of $HTS$
materials - see \cite{Sun1},\cite{Tolpygo}. The plasmon frequency $\omega
_{pl}$ is extracted from the slope $\alpha $ in the linear resistivity $\rho
(T)=\rho _i+\alpha T$, where $(4\pi /\omega _{pl}^2)=\alpha /2\pi \lambda
_{tr}$ and $\lambda _{tr}$ is the transport coupling constant. The realistic
guess for $\lambda _{tr}$ is $\lambda _{tr}\approx 0.3$ \cite{Gurvitch}.
However, it is not correct to replace $\Gamma _s$ by $1/\tau _{tr}$ because
for the latter backward processes dominate. It turns out that in the
presence of strong correlations one has $\Gamma _s=p\cdot 1/\tau _{tr}$,
i.e. $\Gamma _s$ $=2p\pi \lambda _{tr}(\rho _i/\alpha )$ and for/below
optimal doping $\delta \leq 0.2$ it is obtained $p\approx 2$ and $\beta \geq
0.8$ \cite{Zeyher1},\cite{Zeyher2}.

We discuss here three scenarios for the suppression of the measured slope $%
(dT_c/dx)_{x=0}$ ($x=\rho _i/\alpha $), which are compared to the prediction
of the $A-G$ theory \cite{A-G},\cite{Tolpygo} for $d-wave$ pairing with
isotropic impurity scattering.

$1)$ {\bf $d-wave$ pairing ($\langle \Delta (\theta )\rangle _\theta =0$)
and anisotropic scattering ($\beta \neq 0$)}

Based on the above analysis and from eq.(8) one has

\begin{equation}
(dT_c/dx)_{x=0}=-p(1-\beta )\lambda _{tr}\pi ^2/8,  \label{eq.9}
\end{equation}
where $x=\rho _i/\alpha $. The experimental value is $\mid ~dT_c/dx\mid
_{x=0}\approx 0.25$ \cite{Sun1},\cite{Tolpygo},\cite{Chien},\cite{Neu}.
 For $p\approx 2$ and $%
\lambda _{tr}\approx 0.3$\cite{Gurvitch} one obtains $\beta ^{\exp }\approx
0.85$. Such a large experimental value of $\beta $ is in accordance with the
prediction of the theory of strongly correlated electrons\cite{Kulic1},\cite
{Zeyher1},\cite{Zeyher2}. This means that the present theory can \underline{%
consistently} explain the experimental slope $\mid ~dT_c/dx\mid _{x=0}$\cite
{Sun1},\cite{Tolpygo} - see Fig.2.\\[0.9cm]
\epsfysize=4.4in
\hspace*{0.3cm}\epsffile{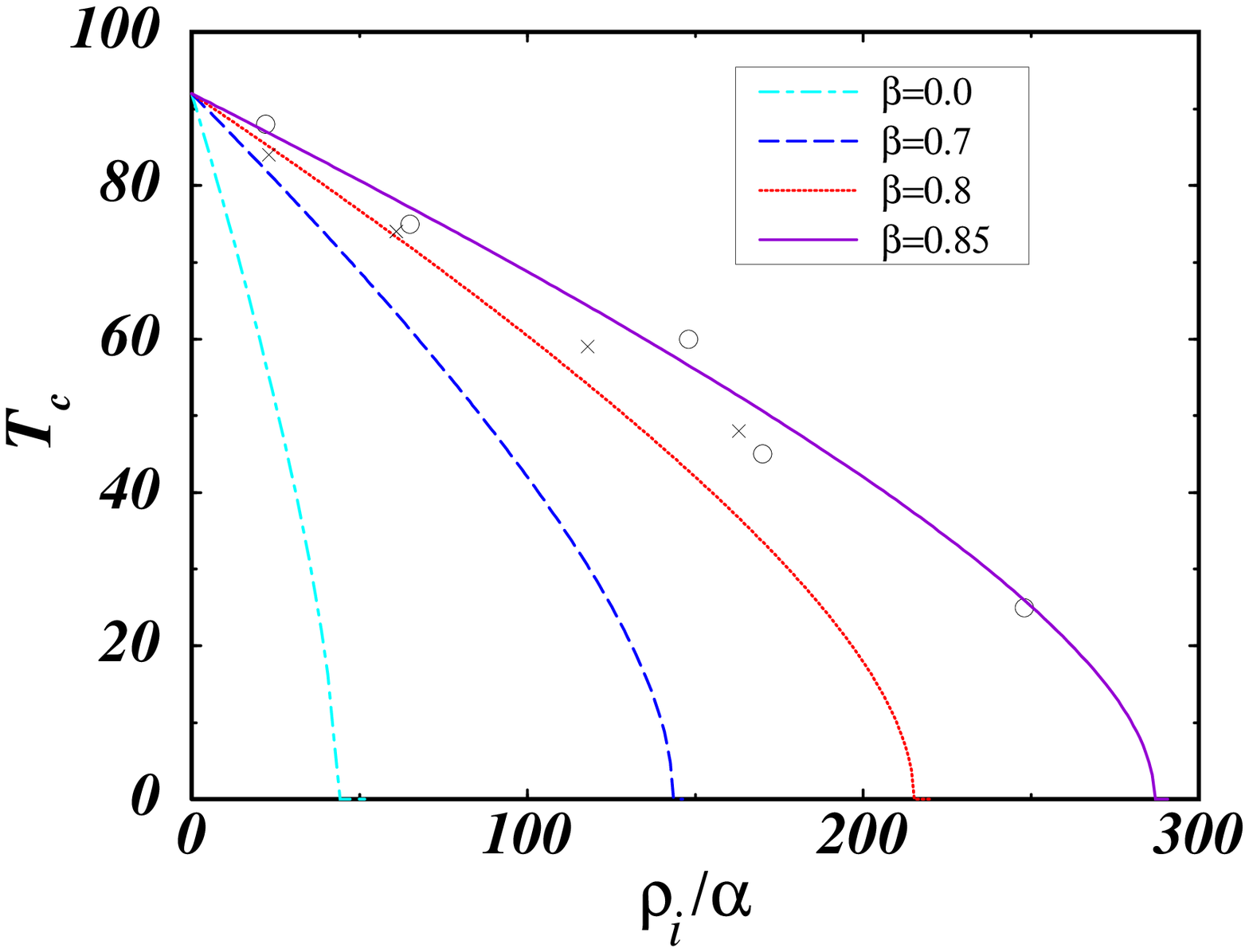}
\vspace*{-5.0cm}
{FIG. 2. The critical temperature $T_c$[K] of $d-wave$ superconductor as a
function of the experimental parameter $\rho _i/\alpha $[K]. The case $\beta =0$
is the prediction of the standard $d-wave$ theory with isotropic scattering.
The experimental data are given by crosses -- 
YBa$_{2}$Cu$_{3-x}$Zn$_{y}$O$_{7-\delta}$ \cite{Chien},
 and circles -- Y$_{1-y}$Pr$_{y}$Ba$_2$Cu$_3$O$_{7-\delta}$  \cite{Neu}.}

$2)$ {\bf Anisotropic pairing ($\langle \Delta (\theta )\rangle _\theta \neq
0$) and isotropic scattering ($\beta =0$)}

Based on eqs.(2)-(5) one obtains eq.(9), where $\beta \rightarrow \beta
_{an}=\langle \Delta (\theta )\rangle _\theta ^2/\langle \Delta ^2(\theta
)\rangle $ -- see \cite{Hohenberg}. For $\mid dT_c/dx\mid _{x=0}\approx 0.25$
one has $\beta _{an}^{ex}\approx 0.85$, i.e. $\langle \Delta (\theta )\rangle
_\theta \approx 0.9\sqrt{\langle \Delta ^2(\theta )\rangle }$. This result
means that a large $s-wave$ component appears in the order parameter, i.e. $%
\Delta (\theta )=\Delta _s+\Delta _dY_d(\theta )$ and $\Delta _d\approx
0.5\Delta _s$. This result is compatible with the experimental observation
of the finite critical current in the $Pb-YBCO$ Josephson contact along the $%
c-axis$\cite{Sun2}, but it is incompatible with $d-wave$ like pairing ($\mid
\langle \Delta (\theta )\rangle _\theta \mid /\Delta _{\max }\ll 1$) found
in many experiments.

$3)$ {\bf Anisotropic pairing ($\langle \Delta (\theta )\rangle _\theta \neq
0$) and anisotropic scattering ($\beta\neq 0$)}

In this case one should set $\beta \rightarrow \beta _{eff}$ in eq.(9),
where $\beta _{eff}=(\beta \Delta _d^2+\Delta _s^2)/(\Delta _d^2+\Delta _s^2)
$. Since $\beta _{eff}^{\exp }\approx 0.85$ one obtains $\Delta _s=2\Delta _d%
\sqrt{1-\beta /0.85}$. This means that finite $\beta $ suppresses the $s-wave$
component in $\Delta ({\theta })$. In our opinion, the realistic situation
for the optimal doping might be described by the case $1)$, or by $3)$ with $%
0.78<\beta <0.85$ and $0<\Delta _s<\Delta _d/3$.

In conclusion, the effect of nonmagnetic impurities on $d-wave$
superconductivity in $HTS$ materials is studied within the \underline{theory
of strong correlations} \cite{Kulic1},\cite{Zeyher1},\cite{Zeyher2} which
predicts: $(i)$ by lowering doping $\delta $ the forward scattering peak is
developed in a nonmagnetic scattering where a large $d-wave$ component is
present in it; $(ii)$ transport properties are renormalized in such a way
that $1/\tau _{tr}=\Gamma _s/p$, where $p\geq 2$ for $\delta <0.2$. Based on
this theory, we have shown that the robustness of $T_c$ in $HTS$ systems in
the presence of defects and impurities is due to: (a) an anisotropic
scattering (due to strong correlations) and (b) the anisotropic pairing with 
$\langle \Delta (\theta )\rangle _\theta =0$ or $\mid \langle \Delta (\theta
)\rangle _\theta \mid /\Delta _{\max }\ll 1$. The theory, which is generally
formulated for the bi-layer model with intra- and inter-layer pairing and
with intra- or inter-layer impurities, shows that the underdoped and
optimally doped systems are robust in the presence of impurities, while the
overdoped systems lose this robustness.

{\bf Acknowledgments:}

One of us (M. L. K.) would like to thank Professors Michael Mehring, Ole K.
Andersen and Lars Hedin for support. We thank Professors Konstantin Efetov
and Vladimir Prigodin for discussions and important comments on the
manuscript. Partial financial support by the Russian State Program
``High-Temperature Superconductivity'', Grant No 95065, and by the Russian
Foundation for Fundamental Research, Grant No 96-02-17527, is acknowledged
by one of us (V.O.)

\end{multicols}

\end{document}